\documentclass[preprint,prd,aps,tightenlines]{revtex4}

\usepackage{amsmath}
\usepackage{verbatim}
\usepackage{graphicx}
\usepackage[usenames]{color}
\usepackage{psfrag}

\def\beq{\begin{equation}}
\def\eeq{\end{equation}}
\def\bea{\setlength\arraycolsep{1.4pt}\begin{eqnarray}}
\def\eea{\end{eqnarray}}
\def\bit{\begin{itemize}}
\def\eit{\end{itemize}}

\begin{document}

\title{Let's talk about varying G}
\author{Adam Moss} \email{adammoss@phas.ubc.ca}
\affiliation{Department of Physics \& Astronomy\\
University of British Columbia,
Vancouver, BC, V6T 1Z1  Canada}
\author{Ali Narimani} \email{anariman@phas.ubc.ca}
\affiliation{Department of Physics \& Astronomy\\
University of British Columbia,
Vancouver, BC, V6T 1Z1  Canada}
\author{Douglas Scott} \email{dscott@phas.ubc.ca}
\affiliation{Department of Physics \& Astronomy\\
University of British Columbia,
Vancouver, BC, V6T 1Z1  Canada}

Essay written for the Gravity Research Foundation 2010 Awards for
Essays on Gravitation

\date{\today}

\begin{abstract}
It is possible that fundamental constants may not be constant at all.  There
is a generally accepted view that one can only talk about variations of
{\it dimensionless\/}
quantities, such as the fine structure constant
$\alpha_{\rm e}\equiv e^2/4\pi\epsilon_0\hbar c$.  However, constraints on the
strength of gravity tend to focus on $G$ itself, which is problematic.
We stress that $G$ needs to be multiplied by the square of a mass, and
hence, for example, one should be constraining
$\alpha_{\rm g}\equiv G m_{\rm p}^2/\hbar c$, where $m_{\rm p}$ is the
proton mass.  Failure to focus on such dimensionless quantities makes it
difficult to interpret the physical dependence of constraints on the
variation of $G$ in many published studies.  A
thought experiment involving talking to observers in another universe
about the values of physical constants may be useful for distinguishing what
is genuinely measurable from what is merely part of our particular system
of units.
\end{abstract}

\maketitle
\noindent
{\em Introduction.---}%
It seems reasonable to suppose that the fundamental constants of nature
could vary with time (or position).  There has been a wide and varied
literature on this topic, from at least the time of the numerological
ponderings of Dirac \cite{dirac1,dirac2} to claims of detection
(and counter-claims
of null results) on the variation of the fine-structure constant over
cosmological time (e.g.~\cite{webb1,webb2,srianand}).
For many researchers interested in
fundamental physics, any variation in such constants could be a clue
for physics beyond the current standard paradigms
(e.g.~\cite{olive,steinhardt}).

There is a long history of debate about whether one can measure the
time-variation of a {\it dimensionful\/} constant.  Although claims to
the contrary
continue to crop up, there is consensus among a long and illustrious line
of physicists that only the variation of {\it dimensionless\/} combinations
of constants can be meaningfully discussed.  Among those pointing this out
are Dicke~\cite{dicke1}, McCrea~\cite{mccrea}, Rees~\cite{rees1},
Jeffreys~\cite{jeffreys}, Hoyle and Narlikar~\cite{hoyle},
Carter~\cite{carter}, Silk~\cite{silk} and Wesson~\cite{wesson}.

The basis for this view is that physical units are quite
arbitrary, and that each individual constant can be removed through a
suitable choice of units
(see e.g.~\cite{duff1,duff2,rich} for a comprehensive discussion).
The simplest example is the speed of light, $c$,
whose variation is now manifestly unmeasurable because of its designation as a
fixed constant relating the definition of time and distance units.
Dimensionless quantities,
on the other hand, are independent of the choice of units and so should admit
the possibility of genuinely observable changes.

There are two distinct contexts in which variable constants are discussed.
The first is to imagine that they are functions of time, $\alpha_{\rm e}(t)$
for example (e.g.~\cite{langacker}).
The second is to imagine different volumes of the Universe, or
different realizations of the Universe, having different constants, related,
perhaps, to ideas of the string landscape (e.g.~\cite{hogan}).
As we discuss below,
even if one considers the idea of the multiverse to be unreasonably
speculative, it may still be helpful to consider how one would communicate
with some alien culture about the possibility of measuring different values
for physical constants.  Conceptualizing how one would have such a
conversation may help focus on what is genuinely measurable.

Much has been written (see e.g.~\cite{Uzan})
about variation in the fine structure constant,
\beq
\alpha_{\rm e} \equiv {e^2\over 4\pi\epsilon_0 \hbar c}
\eeq
(where we have retained SI units in order not to obscure arguments about
dimensions and units). Much of the activity in recent years has been inspired
by claims of measurements or tight constraints on the variability of this
constant with time \cite{murphy,petitjean,kanekar, cmb1, cmb2}.

There have also been many studies aimed at constraining variations of
Newton's gravitational constant, $G$.  $G$ is of course a dimensionful constant,
and hence can no more be measured to vary as can the speed of light or the
charge on the electron. We believe that any study of a variable $G$, should be
restated in a dimensionless manner, as we discuss in the following examples.

\noindent
{\em Dimensionless gravity.---}%
There is a very close analogy between the Coulomb force and classical
gravity, and hence one can define a
`gravitational fine structure constant' :
\beq
\alpha_{\rm g}\equiv {G m_{\rm p}^2\over \hbar c}.
\eeq
This choice of notation goes back at least to Silk in 1977~\cite{silk}.
It explicitly selects the the proton mass as the `gravitational charge',
although any other particle mass can be used instead
(there would then be ratios of $m_{\rm X}/m_{\rm p}$ in
some quantities).  Using current values for the fundamental constants
one obtains
$\alpha_{\rm g}=5.9\times10^{-39}$.  The fact that this is so small
underscores the weakness of gravity compared with electromagnetism.

What the definition of $\alpha_{\rm g}$ means is that
experimental observations should be sensitive only to $G$ multiplied by
the square of the gravitational charge (and normalized by Planck's constant
and the speed of light).
Just as in the case of variation of $\alpha_{\rm e}$, one is {\it not\/}
allowed to ask {\it which\/} of the parts of $\alpha_{\rm g}$ are varying
with time.

Any discussion of the variation of the strength of gravity should start
from this point.  In other words, constraints which appear to have been
placed on the variation of $G$ only make sense if they can be interpreted
as constraints on $G m_{\rm p}^2$.

It should also be noted that just because a quantity is dimensionless does
{\it not\/} imply it is unit-independent. ${\dot G}/G$ (or
$\Delta {G}/G$) suffers from exactly
the same unit-dependence problems as $\dot{G}$.

\noindent
{\em Stars and Galaxies.---}%
A well known example of the use of $\alpha_{\rm g}$ in astrophysics is in the
determination of the mass of a star from first principles:
\beq
M_\ast \sim \alpha_{\rm e}^{3/2} \alpha_{\rm g}^{-3/2}
\left({m_{\rm e}\over m_{\rm p}}\right)^{-3/4} m_{\rm p},
\eeq
or in other words the characteristic size of a star is ${\sim}\,10^{57}$
protons.  This is similar to the Chandrasekhar mass:
\beq
M_{\rm Ch} \sim \alpha_{\rm g}^{-3/2} m_{\rm p} \sim 10^{57} m_{\rm p}.
\eeq
This means that if one were to compare such a mass measured today with the
same mass measured billions of years ago, then one could determine that there
had been a change because a {\it pure number\/} (in this case the number
of nucleons in a star) was different.

An estimate developed later
gives the mass of a galaxy from a cooling time argument
(e.g.\cite{silk,ReesOstriker}:
\beq
M_{\rm gal} \sim \alpha_{\rm e}^5 \alpha_{\rm g}^{-2}
 \left({m_{\rm p} \over m_{\rm e}}\right)^{1/2} m_{\rm p}
 \sim 10^{69} m_{\rm p}.
\eeq
If we wish we can rewrite these in terms of the Planck mass by using
$m_{\rm p}=\alpha_{\rm g}^{1/2}m_{\rm Pl}$.

We can also estimate (see e.g.~\cite{padman})
the characteristic lifetime of a Main Sequence star as:
\beq
t_{\rm gal}\sim \alpha_{\rm e}^{-1} \alpha_{\rm g}^{-1}
\left({m_{\rm e}\over m_{\rm p}}\right)^{-1/2} {\hbar \over m_{\rm e}c^2}.
\eeq
We can rewrite this as a dimensionless number of Planck times, where
\beq
t_{\rm Pl}\equiv \left({\hbar G \over c^5}\right)^{1/2}.
\eeq

\noindent
{\em Aliens and other universes.---}%
What do we really mean when we say that the strength of gravity has changed?
We are usually talking about a change in $\alpha_{\rm g}$ over cosmological
timescales.  However, it is important to keep in mind that there must be
an experimental approach which can measure this variation
{\it relative\/} to something (see \cite{rich}).
Thinking this through carefully can
be particularly important in cosmological studies, since dependence on $G$
(and hence $\alpha_{\rm g}$) can creep in through densities and through
the expansion rate via the Friedmann equation.

Let us consider how we might, in principle, compare notes with observers
in some different universe, with potentially different physical
parameters (e.g.~\cite{adams}).  Since their units are likely to be completely
different, we would focus on dimensionless quantities, which for gravity
means $\alpha_{\rm g}$.  In our discussion with these alien observers we would
easily be able to focus on the crucial dimensionless parts of physics.
We would also be able to say something like: `you know that mass where
the Schwarzschild radius is equal to the Compton wavelength?  Well, we call
that the Planck mass, and let's agree to use that as a fundamental scale'.
We would continue the dialogue until we had also agreed on the
definition of the other Planck units.  Then if a particular experiment
constrains a length, say, we know to describe the result as a
certain number of Planck lengths, with any difference being ascribed to the
required dependence on dimensionless `constants'.

\noindent
{\em Recombination.---}%
To illustrate this, take the example of cosmological recombination
(see e.g.~\cite{SSS}).  The
variation of fractional ionization of hydrogen with redshift can be expressed
in its simplest form through the equation:
\beq
{dx\over dz} \sim {x^2n\alpha_{\rm rec}\over H(z)(1+z)} \, ,
\eeq
where $H(z)$ is the Hubble expansion rate, $n$ is the number density of
hydrogen atoms (baryons) and $\alpha_{\rm rec}$ is the recombination rate
(unfortunately using the same Greek letter as our dimensionless constants,
but here with units $m^3\,{\rm s}^{-1}$).
The solution of the recombination equation is
observable and dimensionless, and hence
must depend only on dimensionless physical quantities.
By considering the physical dependence within $\alpha_{\rm rec}$,
using $H^2\sim G\rho$ and writing the ratio of baryon-to-photon number
densities as $R_{\rm B\gamma}$,
we find the dependency of the equation reduces to
\beq
\alpha_{\rm e}^3 \alpha_{\rm g}^{-1}
 \left({m_{\rm e}\over m_{\rm p}}\right)^{-3/2}
 R_{\rm B\gamma}^{-1/2} \left({kT \over E_{\rm Pl}}\right),
\eeq
where $E_{\rm Pl}$ is the Planck energy, $E_{\rm Pl}\equiv\sqrt{\hbar c^5/G}$.

This dependence defines the thickness of the last-scattering surface,
which is a large part of the effect of variable $G$ on Cosmic Microwave
Background anisotropies (see e.g.~\cite{galli,umezu}).

\noindent
{\em Big Bang Nucleosynthesis.---}%
Another example, which has also been used to constrain the variation of
$G$ is the abundance of the light elements through nucleosynthesis in
the early Universe (e.g.~\cite{Iocco} and references therein).

The crucial physics is determined by the condition for freeze-out of the
weak interactions.  Setting a particular form for the weak reaction rate
equal to the Hubble rate leads to
\beq
\left( {kT\over E_{\rm Pl}}\right)^3 \sim \alpha_{\rm w}^{-4} \alpha_{\rm g}^2
\left({m_{\rm W}\over m_{\rm p}}\right)^4,
\eeq
where $\alpha_{\rm w}$ is the dimensionless coupling constant for the
weak interaction and $m_{\rm W}$ is the W-boson mass (there are other
formulations in terms of the vacuum expectation value of the Higgs field,
but we need not concern ourselves here with details of electroweak physics).
The main point is that this is different from the dependence which some authors
have claimed to be constraining.

\noindent
{\em Conclusions.---}%
We have not explored all examples in the literature of constraints on
$G$, and we have not exhaustively assessed each and every paper.  However,
it is clear that at least some of the published discussions involving
$G$ rather than $\alpha_{\rm g}$ are in fact constraints on a different
combination of parameters than asserted by the authors.

The possibility of variation of the `constants' of nature continues to
intrigue physicists.  Dirac's 1937 paper on `The Cosmological Constants'
has been cited more than 500 times, and there are numerous other papers
claiming to constrain variations of $c$, $e$, $h$ or $G$.  The situation
regarding the strength of electromagnetism has become quite clear, with
papers discussing anything other than the dimensionless combination
$\alpha_{\rm e}$ now being quite rare.  However, the literature is still
replete with discussions of limits on $G/{\dot G}$ using solar system
tests, stellar evolution, light element abundances, microwave anisotropies,
or other tests.  We believe that cosmologists should get their gravitational
house in order, and speak only of $\alpha_{\rm g}$.  Imagining how one would
talk meaningfully about physical constants with observers in a
different universe may, ironically, be helpful in clarifying what is being
measured in our own back yard.

\noindent
{\em Acknowledgments.---}%
This research was supported by the Natural Sciences and Engineering 
Research Council of Canada.  We thank the many colleagues with whom we have
had fruitful discussions on this topic, in particular Jeremy Heyl and Jim
Zibin.

\end{document}